\documentclass[pra, twocolumn]{revtex4}
\usepackage{graphicx}
\usepackage{amssymb}
\usepackage{epstopdf}
\usepackage{color}
\usepackage{subfigure}

\newcommand{\tr}{{\textrm{tr}}}

\begin{document}

\title[Entanglement in the Bose-Einstein condensate phase transition.]{Entanglement in the Bose-Einstein condensate phase transition}

\author{Libby Heaney}

\address{The School of Physics and Astronomy, University of Leeds, Leeds LS2 9JT, UK}

\begin{abstract} 
We determine the behaviour of entanglement between regions of space in a Bose gas of fixed particle number around the critical temperature condensation.  Long-range correlations develop in the Bose-Einstein condensate (BEC) phase transition and the aim here is to find out whether spatial coherence alone implies entanglement.  We use a purity measure of entanglement to derive an entanglement witness that detects entanglement between two regions of space in the BEC.  It is shown that spatial coherence between the two regions is necessary for entanglement with coherence and entanglement becoming equivalent only when the regions occupy the entire confining volume of the gas.  The probabilities for Bosons to occupy the regions is the only other parameter that influences the amount of entanglement.  We  calculate explicitly the amount of entanglement between two regions for a cigar-shaped harmonic potential and find that it increases with decreasing temperature.  A second entanglement witness is derived for entanglement between three regions of space, where again spatial coherence between each pair of regions is precursor to entanglement.  It is shown that the state with the maximum amount of entanglement occurs when the probabilities for the Bosons to occupy the three regions are equal.  
%
\end{abstract}
\maketitle

\section{Introduction}

The presence of entanglement in phase transitions has attracted considerable interest over the past few years.  Understandably, quantum phase transitions were the natural place where one expected to find the non-local correlations of entanglement \cite{Osterloh02,Osborne02,Vidal03}.  A quantum phase transition is the qualitative change in the ground state of a many-body system brought on at zero temperature by the change of an external parameter or coupling constant.  At the quantum critical point long-range correlations develop in the ground state and an extra quantity known as the quantum order parameter is required to fully describe the system.  
The motivation for investigating the entanglement in quantum phase transitions lies in the fact that the ground states of most condensed matter many-bodied systems occupy huge Hilbert spaces and are difficult to construct.  While one may not have full knowledge of such a ground state, one can still explore how the entanglement changes close to quantum critical points, thus accessing some aspects of the ground state's behaviour in critical regimes.  Moreover, it has been argued that entanglement can provide a physical picture and a deeper understanding of renormalisation methods \cite{Osborne02a}, quantum phase transitions \cite{Osborne02} and their associated quantum order parameters \cite{Brandao05}.  

Most research studying entanglement in phase transitions concerns discrete lattice type systems and to date hardly any research has focused on entanglement in the Bose-Einstein condensate (BEC) phase transition.  Initially, Simon \cite{Simon02} demonstrated that any two regions of a uniform zero-temperature condensate of finite particle number are entangled.  A recent paper \cite{Heaney07} demonstrated that entanglement exists only below the critical temperature for condensation, but a precise relationship between entanglement and the long-range order was not specified.  The lack of research into entanglement in the BEC phase transition stems from the fact that  there is no entanglement between individual particles in a non-interacting BEC.  Usually entanglement is thought to exist between individual particles or their internal degrees of freedom, but in this case one has to instead consider entanglement between regions of space occupied by particles.  

The ground state of a non-interacting Bose gases is already well understood, which means that we do not need to use entanglement to characterise it's behaviour.  So why, in this case, is it useful to study entanglement in the BEC phase transition?  While the BEC phase transition is classical, i.e. it occurs by varying the temperature,  quantum effects develop on a macroscopic scale without the presence of interactions.   This makes a non-interacting BEC a remarkably simple platform for establishing concrete connections between entanglement and the long-range correlations that develop in phase transitions.  If we can fully understand how entanglement emerges in the simple setting of the BEC phase transition, we can use these results as a guide for understanding how entanglement and long-range correlations are related in more complicated systems.  For instance, Osborne and Nielsen \cite{Osborne02, Osborne02a} conjecture that entanglement is responsible for the long-range correlations that develop in the transverse Ising and XY models at the quantum critical point and that at such critical points the system is maximally, or {\it critically}, entangled.   

In this paper we will develop a thorough understanding of entanglement in the BEC phase transition and as a consequence we will be able to provide evidence for and against such statements.  We will see that spatial coherence of a Bose gas is necessary for entanglement between regions of space.  The probabilities for the Bosons to occupy the regions is the other crucial parameter affecting the entanglement.  So while there may be spatial coherence between two points in space, these points will not be entangled because there is almost a zero probability of Bosons populating the points.  In this way we will be able to understand what portion of the long-range correlations are actually entanglement.  

The remainder of this paper is organised as follows.  We will first discuss the concept of long-range order and spatial coherence of a BEC in section \ref{BECPhaseTrans},  followed by brief discussions of spatial entanglement in section \ref{spatialent} and an overview of an entanglement measure based on the purities of states in section \ref{purity}.  In section \ref{calculationsec} we will calculate the entanglement between two regions of space around the critical temperature for condensation.  An inequality relating the entanglement to spatial coherence will be derived in section \ref{entLROsec}.  We will compare our theoretical results for entanglement  to recent experimental results measuring the spatial coherence of a BEC in section \ref{expresultssec}.  To fully characterise entanglement in the BEC phase transition we will consider multi-mode entanglement in section  \ref{multientsec}.  Finally in \ref{concludesec}, we will conclude by discussing the results of this paper in the context of previous results concerning the role of entanglement in quantum phase transitions of discrete systems.  
 
 \subsection{The BEC phase transition}\label{BECPhaseTrans}
 
A simple but important example of a phase transition is the formation of a BEC.  As a Bose gas is cooled close to absolute zero, it passes a critical temperature where the chemical potential becomes comparable to the ground state energy causing a macroscopic number of Bosons to  collapse into the ground state of the confining potential.  As a result, all of the Bosons behave as a single wave and the phase of the condensate becomes locked.  The wavefunction of the condensate then plays the role of the order parameter.  The appearance of an order parameter signifies that long-range correlations have developed between two distant points of a condensate.  

Such long-range correlations were described mathematically by Penrose and Onsager \cite{Penrose56} in terms of the long-range order (LRO). LRO is present in a condensate when the one-body reduced density matrix 
\begin{equation}
\rho_1(\vec{r},\vec{r}^{\, \prime})=\tr[\Psi^{\dagger}(\vec{r})\Psi(\vec{r}^{\, \prime})\rho],
\end{equation}
is finite as $|\vec{r}-\vec{r}^{\, \prime}|\rightarrow\infty$, where $\Psi^{\dagger}(\vec{r})$ and $\Psi(\vec{r})$ are the creation and annihilation operators of particles at point $\vec{r}$.  This definition holds for interacting and non-interacting gases, but because $\rho_1(\vec{r},\vec{r}^{\, \prime})$ tends to the ground state density, $\rho_1(\vec{r},\vec{r}^{\, \prime})\rightarrow n_0$, for $|\vec{r}-\vec{r}^{\, \prime}|\rightarrow\infty$, its value is diminished by particle interactions.  The one-bodied reduced density matrix, $\rho_1(\vec{r},\vec{r}\,')$, measures the presence of off-diagonal matrix elements in the state and determines the quantum coherences between two points.  The aim of this paper is to investigate the how the spatial coherence, described by $\rho_1(\vec{r},\vec{r}\,')$, are related to entanglement.

Let us briefly consider whether the one-bodied reduced density matrix, $\rho(\vec{r},\vec{r}\,')$, is a good measure of entanglement. If we imagine one particle distributed at two points, $\vec{r}$ and $\vec{r}\,'$, with equal probability, the state of the two points occupied by particles would be $|\psi_1\rangle=1/\sqrt{2}(|10\rangle+|01\rangle)$, where $|01\rangle=|0\rangle_A\otimes|1\rangle_B$.  A two particle state is $|\psi_2\rangle=1/\sqrt{3}(|20\rangle+|11\rangle+|02\rangle)$ and a $N$ particle state would be $|\psi_N\rangle= 1/\sqrt{N}(|N0\rangle+|N-1,\, 1\rangle+ \dots+|1,\,N-1\rangle+|0N\rangle)$.  The $N$ particle state, $|\psi_N\rangle$, clearly contains more entanglement than the one or two particle state, but the normalised one-bodied reduced density matrix is identical for all three cases.
So while the one-body density matrix can detect entanglement, it cannot differentiate between entanglement formed by one or N particles occupying the regions.

\subsection{Entanglement between spatial modes}\label{spatialent}

What does it mean for two regions of space to be entangled?  We know from the simple form of  Hamiltonian of a non-interacting BEC, $\hat{H}=\sum_{\vec{k}} E_{\vec{k}} \hat{a}_{\vec{k}}^{\dagger}\hat{a}_{\vec{k}}^{}$,  that individual particles are decoupled  in momentum representation and are therefore not entangled to one another.  However, one could also view this Hamiltonian in position representation. Here the operators, $a^{\dagger}_{\vec{k}}, a_{\vec{k}}$, that create and destroy particles in mode $k$, transform as $\hat{a}_{\vec{k}}^{\dagger}=\int dx \phi_{\vec{k}}(x) \hat{\psi}^{\dagger} (x)$, where $\phi_{\vec{k}}(x)$ is the eigenvector for mode $\vec{k}$ and $\hat{\psi}^{\dagger}(x)$ creates a particle at position $x$.  In position representation, one can split the integral into multiple regions so that there are pairwise couplings between all regions of space, $\hat{H}=\sum_{\vec{k}}E_{\vec{k}}\sum_{i,j}\int_id\vec{r}\,\int_jd\vec{r}'\,\phi_{\vec{k}}(\vec{r})\phi_{\vec{k}}^*(\vec{r}')\hat{\psi}^{\dagger}(\vec{r})\hat{\psi}(\vec{r}')$. This is in fact what leads to entanglement between regions of space.  

Spatial entanglement therefore takes the form of non-local particle number correlations between regions of space or {\it spatial modes}. The spatial modes are entangled and not the particles themselves. In order to ensure the correct Hilbert space structure one must work with second quantised modes and not in the first quantised picture.  In second quantisation the Hilbert space of three spatial modes occupied by Bosons have a tensor product structure.  One is then able to trace out each mode to obtain a state of just two regions of the gas. For a more detailed discussion of this and other aspects of spatial entanglement see references \cite{ Heaney07,Anders06, Heaney06, Cunha07, Dunningham07}.

\subsection{Entanglement of continuous variable states using purity measurements}\label{purity}

In this paper  we will use a measure of entanglement based upon the purities of the global and reduced states of the spatial modes.  
The purity, $\mu$, measures the degree of information contained in a quantum state, $\rho$, and is defined as $\mu = \textrm{tr}[\rho^2]$.   The purity is related to the mixedness, $S_L(\rho)$, of the state by $S_L(\rho)=[N/(N-1)](1-\mu)$, where $N$ is the number of modes.  The purity ranges from  one for pure states to $1/D$ for fully mixed states, $1/D\leq \mu \leq 1$, where $D=\textrm{dim}\,\mathcal{H}$, is the dimension of the Hilbert space.  For a state, $\rho_{AB}$, of two modes $A$ and $B$, the global purity is defined as $\mu=\tr[\rho_{AB}^2]$ and the reduced purity of the $i$th mode is $\mu_i=\tr[\rho_i^2]$, where $\rho_i$ is the reduced density operator of mode $i=A$ or $B$.

It is well known that the concepts of entanglement and the information encoded in a quantum states are closely entwined. For instance, in pure states, entanglement and the lack of information, or {\it mixedness}, of the reduced state of a subsystem are equivalent.  A maximally entangled pure state will have maximally mixed reduced states. Thermal states on the other hand have a mixed global state.  Entanglement therefore exists in mixed states if the global mixedness of a quantum state is less than some function of the mixedness of the reduced states of the subsystems. 

An entanglement measure in terms of global and reduced purities for continuous variable Gaussian states \cite{Adesso04} is
 \begin{equation}
 \frac{\mu_A\mu_B}{\sqrt{\mu_A^2+\mu_B^2-\mu_A^2\mu_B^2}}<\mu\leq\frac{\mu_A\mu_B}{\mu_A\mu_B+|\mu_A-\mu_B|}.
 \end{equation}
When this inequality is satisfied, there is entanglement between modes $A$ and $B$.  The upper bound represents a physicality condition on the quantum state, $\rho_{AB}$.  Just below the lower bound there is a small region where entangled and separable states coexist. We shall see shortly that for systems with a large number of non-interacting particles, the region where separable and entangled states both exist goes to zero. 

 In what follows we shall choose the spatial modes so that the reduced purities of $A$ and $B$ are equal $\mu_A=\mu_B=\mu_r$.  The entanglement criterion then reduces to
\begin{equation}
 \label{purityrelation}
\frac{\mu_r}{\sqrt{2-\mu_r^2}}<\mu\leq1, 
\end{equation}
and will be used shortly to derive our entanglement witness.

\section{Entanglement between two spatial modes around the critical temperature for condensation}\label{calculationsec}

We will now calculate the entanglement between two spatial modes in a BEC.  Let us consider a non-interacting Bose gas of fixed particle number, $N$, at temperature $T$, placed in some confining volume $V(\vec{r})$, where the particles occupy energy levels determined by the eigenfunction $\phi_{\vec{k}}(\vec{r})$.  The results that we present here are for any confining potential; we will present a specific example in section \ref{expresultssec}.    The purity measure, (\ref{purityrelation}), requires knowledge of  the states of one and two spatial modes, $\rho_{A}$ and $\rho_{AB}$ respectively, where regions $A$ and $B$ do not together occupy the entire volume $V(\vec{r})$. To arrive at, $\rho_{A}$ and $\rho_{AB}$,  the state of the gas is mathematically divided into three regions $A, B$ and $C$, as this allows us to speak about spatial modes. In reality the gas remains unchanged as we are not considering physical partitions at all.  Region $C$ will then be traced out leaving the state, $\rho_{AB}=\textrm{tr}_{C}[\rho_{ABC}]$, of regions $A$ and $B$ that may be entangled.

To find the state, $\rho_{AB}$, of one particle, we first create a particle in mode $\vec{k}$ using the creation operator $\hat{a}_{\vec{k}}^\dagger$ as $|1_{\vec{k}}\rangle=\hat{a}_{\vec{k}}^{\dagger}\,|0\rangle$.  But, as entanglement does not exist in momentum representation, we transform into position representation as $|1_{\vec{k}}\rangle=\int d\vec{r}\,\phi_{\vec{k}}(\vec{r})\hat{\psi}^{\dagger}(\vec{r})|0\rangle$, where the operator $\hat{\psi}^{\dagger}(\vec{r})$ creates a particle at position $\vec{r}$.  To form the three spatial modes $A, B$ and $C$, the integral in $|1_{\vec{k}}\rangle$ is split into the three parts  giving $|1_{\vec{k}}\rangle=(\int_A d\vec{r}\,\phi_{\vec{k}}(\vec{r})\hat{\psi}^{\dagger}(\vec{r})+\int_B d\vec{r}\,\phi_{\vec{k}}(\vec{r})\hat{\psi}^{\dagger}(\vec{r})+\int_C d\vec{r}\,\phi_{\vec{k}}(\vec{r})\hat{\psi}^{\dagger}(\vec{r}))|0\rangle$.  The state of a single particle is  $\rho=\sum_{\vec{k}}P_{\vec{k}}|1_{\vec{k}}\rangle\langle1_{\vec{k}}|$, where each momentum mode is weighted by the probability of finding it in that mode, $P_{\vec{k}}=n_{\vec{k}}/N$, where $\sum_{\vec{k}}P_{\vec{k}}=1$ and $n_{\vec{k}}$ is the Bose distribution \cite{Pitaevskii03}.  

So far we have found the state of one particle expressed in terms of three spatial modes.  To arrive at $\rho_{AB}$, region $C$ is traced over.  We can perform the trace on each momentum mode separately, which results in the state $\rho_{AB}^{(\vec{k})}=\tr_C[\rho^{(\vec{k})}]$ for mode $\vec{k}$, where $\rho^{(\vec{k})}=|1_{\vec{k}}\rangle\langle1_{\vec{k}}|$.  Explicitly $\rho_{AB}^{(k)}$ is
\begin{eqnarray}
\rho_{AB}^{(k)}&=&\sum_{i,j=A,B}\int_id\vec{r}\,\int_j d\vec{r}'\, \phi_k(\vec{r})\phi_k^*(\vec{r}')\hat{\psi}^\dagger(\vec{r})|00\rangle\langle00|\hat{\psi}(\vec{r}')\nonumber\\
&&+\int_Cd\vec{r}\,|\phi_k(\vec{r})|^2|00\rangle\langle00|,\quad\quad\quad\quad
\end{eqnarray}
where the vector $|00\rangle=|0\rangle_A\otimes|0\rangle_B$ is the vacuum for regions $A$ and $B$.  The final state, $\rho_{AB}$, is therefore $\rho_{AB}=\sum_{\vec{k}}P_{\vec{k}}\rho_{AB}^{(k)}$ and likewise for $\rho_A$.  Let us take a closer look at how one performs the trace in this situation.  The operators $\hat{\psi}^{\dagger}(\vec{r}),\,\hat{\psi}(\vec{r})$ act on point $\vec{r}$ which is located within one of the three spatial modes.  The trace of the $i$th spatial mode acts only on operators whose position vector $\vec{r}$ is located in mode $i$.  One can write the trace of the $i$th mode as $\textrm{tr}_{i}[...] = \sum_{n=0,1}\int_i d\vec{r}\langle 0 |(\hat{\psi}(r))^n\,...\,(\hat{\psi}^{\dagger}(\vec{r}))^n|0\rangle$.  

As we are considering a gas of non-interacting Bosons with Hamiltonian $\hat{H}=\sum_{\vec{k}}E_{\vec{k}}\hat{a}_{\vec{k}}^{\dagger}\hat{a}_{\vec{k}}$, it is straightforward to generalise to the case of $N$ particles.  We can write the $N$ particle state in terms of the single particle state as $\rho^{(N)}={\rho}^{\otimes N}$.  
To prove that this is equivalent to the state for BECs found in text books consider one Boson in first quantisation. One can place $N$ particles into this state, ensure it is correctly symmetrised and then move into second quantisation.  Equally one could place $N$ particles in the first quantised state and move directly into second quantisation, which results in the state above.
For $N$ particles, the entanglement criterion (\ref{purityrelation}) can be expressed differently in terms of the purities for the single particle density matrices, $\mu=\textrm{tr}[{\rho_{AB}}^2]$ and $\mu_r=\textrm{tr}[{\rho_A}^2]$.  The purity, $\mu_N$ of $N$ non-interacting particles is $\mu_N=\tr[(\rho^{\otimes N})^2]=\mu^N$, where $\mu$ is the single particle purity.  Using this relationship the inequality for entanglement becomes $\mu-\mu_r>0$ for large $N$.  If the global purity is larger than the reduced purity, the two spatial modes $A$ and $B$ are entangled.  Indeed the upper bound, $\mu\leq1$, is always satisfied if the state, $\rho$, is physical to begin with and for large $N$ ($N>1000$), separable and entangled states never coexist.

The global and reduced purities for a single particle can now be evaluated.    The global purity, $\mu=\tr[{\rho_{AB}}^2]$, is
\begin{eqnarray}
\label{globalpurity}
\mu&=&\sum_{i=A,B,C}p_i^2 + 2\int_Ad\vec{r}\,\int_B d\vec{r}\,'\, \rho^2(\vec{r},\vec{r}'),
\end{eqnarray}
where $p_i=\sum_{\vec{k}}P_{\vec{k}}\int_i d\vec{r}\,|\phi_{\vec{k}}(\vec{r})|^2$ is the probability for occupying the $i$th spatial mode and $\rho(\vec{r},\vec{r}')=\sum_{\vec{k}}P_{\vec{k}}\,\phi_{\vec{k}}(\vec{r})\phi_{\vec{k}}^*(\vec{r}')$. It is immediately clear that the state $\rho_{AB}$ is not pure, as tracing out region $C$ destroys the coherences between it and regions $A$ and $B$.   Equally, the reduced purity for the state of one region contains no coherence terms and for region $A$ reads 
\begin{eqnarray}
\mu_A=\tr[\rho_{A}^2]=(p_B + p_C)^2 + p_A^2= p(p-1)+\frac{1}{2} = \mu_r
\end{eqnarray}
Note that for simplicity the reduced purities for $A$ and $B$ are chosen to be equal, $\mu_A=\mu_B=\mu_r$.

Using the entanglement criterion, $\mu-\mu_r>0$, we can write down a quantity, $\mathcal{E}$, that will be positive when there is entanglement between spatial modes $A$ and $B$ and is negative when the two modes are separable.  The quantity $\mathcal{E}$ can therefore be thought of as an entanglement witness 
and is given by
\begin{eqnarray}
\label{entinequality}
\mathcal{E}= -p(1-2p)+\int_Ad\vec{r}\,\int_Bd\vec{r}'\, \rho^2(\vec{r},\vec{r}').
\end{eqnarray}
When $\mathcal{E}>0$, the two regions $A$ and $B$ are entangled. Eq. (\ref{entinequality}) is a general expression for non-interacting Bose gases as one can consider a gas with any normalised momentum distribution $P_{\vec{k}}$, trapped in any geometry in any dimension.  The second term, or {\it coherence} term, in $\mathcal{E}$ measures the spatial coherence between regions $A$ and $B$.   The first term in $\mathcal{E}$ is parameterised by the probability, $p$, to occupy region $A$ or $B$ and takes into account the fact that while there may be coherences between two regions of any size, there is not necessarily entanglement.  For entanglement to exist the purity of the global state has to be greater than the purity of the reduced state. 

If the regions $A$ and $B$ are half of the volume of the gas each, the probability to be in each region is  $p=1/2$ and the second term in $\mathcal{E}$ is zero.  In this case entanglement is equivalent to spatial coherence between the two regions.  This is particularly interesting as a small amount of coherence, and hence entanglement, will remain between two such regions above the condensation temperature. On the other hand, if we choose two very small regions while there will be spatial coherence between them, there will be a relatively low probability for the Bosons to occupy them and there will be no entanglement.  We will plot $\mathcal{E}$ for a specific system in section \ref{expresultssec}. 

\section{The relationship between entanglement and spatial coherence in a BEC}\label{entLROsec}

The normalised one-body reduced density matrix, $\tilde{\rho}_1(\vec{r},\vec{r}')=\sum_{\vec{k}}(n_{\vec{k}}/N)\,\phi_{\vec{k}}(\vec{r})\phi^*_{\vec{k}}(\vec{r}')$ is identical to  $\rho(\vec{r},\vec{r}')=\sum_{\vec{k}}P_{\vec{k}}\,\phi_{\vec{k}}(\vec{r})\phi_{\vec{k}}^*(\vec{r}')$ used in the expression for entanglement $\mathcal{E}$, Eq. (\ref{entinequality}).  The entanglement witness, $\mathcal{E}$, is already expressed in the language of LRO and spatial coherence.  It is evident that spatial coherence is necessary for entanglement between regions $A$ and $B$,  as the second term in $\mathcal{E}$ is, of course, always positive.  

Rather than looking at the coherences between two points, it may be more appropriate to define a quantity, $O_D=\textrm{tr}[\hat{\psi}^{\dagger}_A\hat{\psi}_B\,\rho]$ that measures the coherence between the regions $A$ and $B$.   The operators $\hat{\psi}^{\dagger}(\vec{r})$ and $\hat{\psi}(\vec{r})$ are averaged over a detector profile $g(\vec{r})$ in the regions, resulting in $\hat{\psi}^{\dagger}_A=\int_A d\vec{r} \,g(\vec{r})\hat{\psi}^{\dagger}(\vec{r})$ and  $\hat{\psi}_B=\int_B d\vec{r} \,g^*(\vec{r})\hat{\psi}(\vec{r})$ that create and destroy particles in regions $A$ or $B$.   The quantity $O_D$ is given by
\begin{equation}
\label{OD}
O_D = \textrm{tr}[\hat{\psi}^{\dagger}_A\hat{\psi}_B\rho] = \int_A d\vec{r}\,\int_B d\vec{r}\,' g(\vec{r})g^*(\vec{r}\,')\rho(\vec{r},\vec{r}\,').
\end{equation}

From elementary statistics we know from the positivity of the variance, $\sigma^2=\langle X^2\rangle-\langle X\rangle^2$, that  the expectation value of some quantity squared  is greater than the square of the expectation value of the quantity, $X$, namely $\langle X^2 \rangle\geq \langle X\rangle^2$.  Thus, the quantity $O_D$ squared is always less than the first term in $\mathcal{E}$- the expectation value of the one-body reduced density matrix squared,
\begin{equation}
\int_Ad\vec{r}\,\int_Bd\vec{r}\,'\tilde{\rho}_1^2(\vec{r},\vec{r}\,')\geq O_D^2.
\end{equation}
The left hand side of this inequality is the coherence term in the entanglement witness, $\mathcal{E}$, which is always greater than or equal to the newly defined coherence term $O_D$.  Using this fact we can define a new quantity $\mathcal{E}_{O_D}$ that also acts as an entanglement witness.  
\begin{equation}
\label{offdiaginequality}
\mathcal{E}_{O_D} = O_D^2 - p(1-2p).\quad\quad
\end{equation}
If the new quantity, $\mathcal{E}_{O_D}$, is greater than zero, $\mathcal{E}_{O_D}>0$, the previous entanglement inequality, $\mathcal{E}>0$, is automatically satisfied and there is entanglement between regions $A$ and $B$.  It is worth expressing the entanglement witness in terms of $O_D$ as this quantity only depends upon the one-body density matrix, $\rho(\vec{r},\vec{r}\,')$, rather than its square, and is straightforward to measure experimentally, see for instance \cite{Haensch02}.


\section{A specific example: BEC in an harmonic trap and comparisons with experiment}\label{expresultssec}
\begin{figure}[t]\centering
{\includegraphics[width=0.5\textwidth]{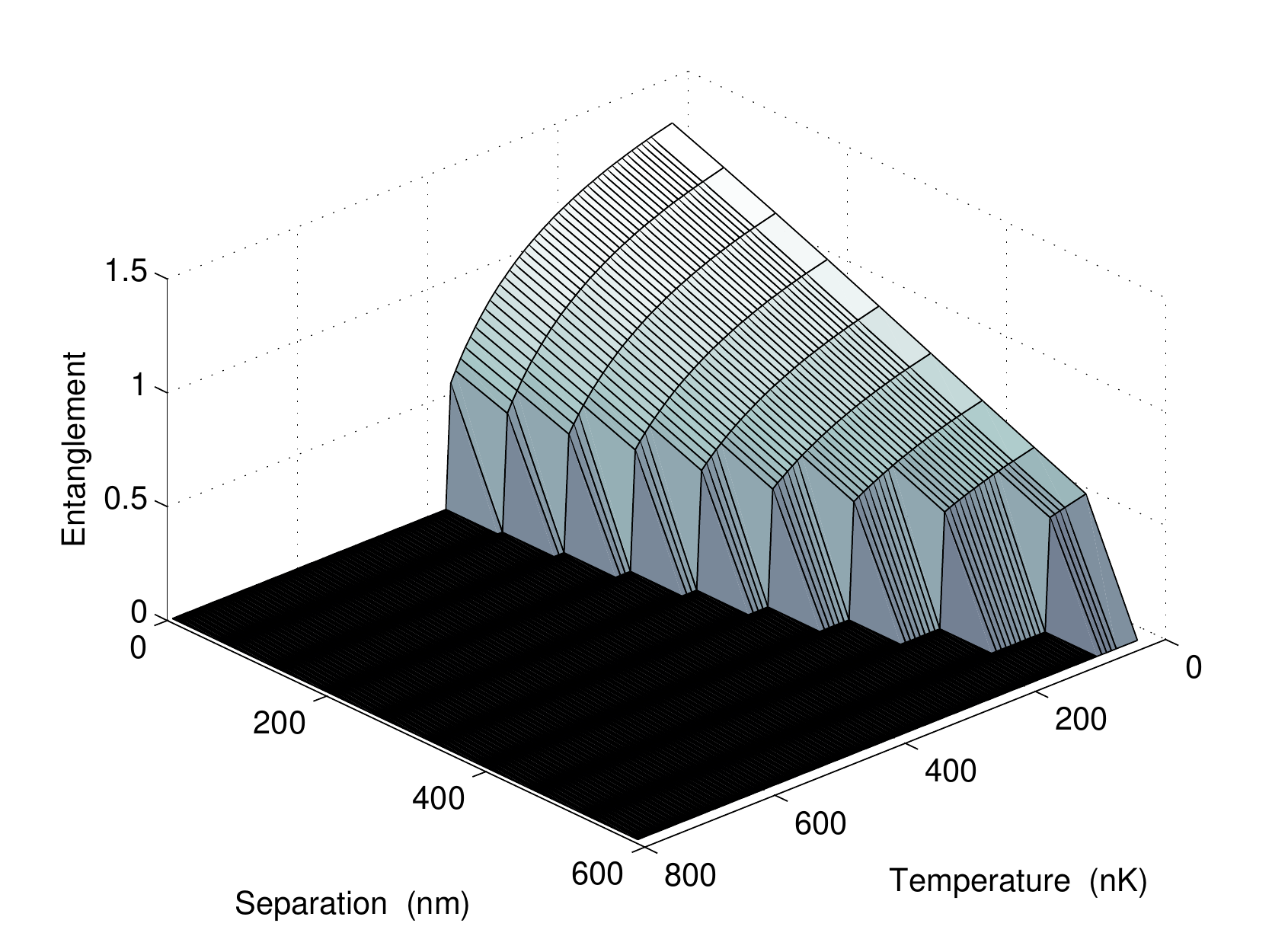}
\caption{ The amount of entanglement as measured by the average logarithmic negativity \cite{Adesso04} between two spatial modes in a harmonically trapped non-interacting BEC of fixed particle number above and below the transition temperature for condensation ($T_C = 440$ nK). See text for the trapping parameters. The amount of entanglement decreases with increasing temperature of the gas.  Due to the Gaussian ground state of the harmonic oscillator the amount of entanglement also decreases as the two regions move further apart. }\label{fig:Fig6}}
\end{figure}

To bring clarity to the results presented in the previous sections we now present a specific example.  We will consider a typical experimental set up, namely a BEC confined in a harmonic cigar shaped trap.  The exact trapping parameters that we use here have been taken from an experimental paper that measures the spatial coherence of a BEC \cite{Haensch02} as this will allow us to make direct comparisons with experiment.   In this case the transition temperature for condensation is $T_C \approx 440$nK, the trapping frequencies are $\omega_{ax}=2\pi\times 13$ Hz and $\omega_{per}=2\pi\times 140$ and the number of Bosons in the gas  is $N\approx4\times10^6$. We take a one-dimensional slice through the trap, define regions $A$ and $B$ of size $L_A=L_B=350$nm and trace out the rest of the gas. The amount of entanglement as determined by (\ref{entinequality}) and quantified by the average logarithmic negativity \cite{Adesso04}  can be seen in Fig. \ref{fig:Fig6}.
Entanglement does not exist above the critical temperature for condensation and the amount of entanglement between the two regions increases as the temperature decreases.  There is no entanglement between the regions when they are separated by more than $\Delta z=600$nm.

However the harmonic nature of the trap hides some of the interesting features of the entanglement.  Therefore let us review the results presented in Simon \cite{Simon02}.  A BEC was considered at zero temperature, where the Bosons occupied all points with equal probability.  The same amount of entanglement was found to exist between two regions of a given size at any separation, i.e. two regions that were touching or two regions that were very far apart have the same amount of entanglement between them.  This is because in position representation there are pairwise couplings between all regions of space (see section \ref{spatialent}) and in \cite{Simon02} the probability to occupy any region of space was equal.  Here entanglement decreases with increasing region separation due to the Gaussian ground state wavefunction.  As the regions move further apart the probability for Bosons to occupy the regions drops, and the state of regions $A$ and $B$ becomes increasingly mixed.

\begin{figure}[t]\centering
{\includegraphics[width=0.45\textwidth ]{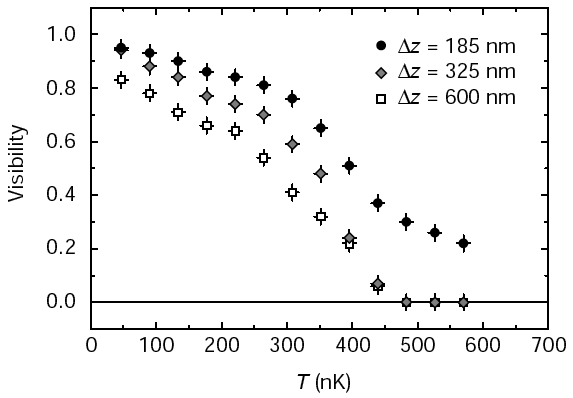}
\caption{ Spatial coherence of a BEC above and below the transition temperature $T_C$ as measured by Bloch in {\it et al.} \cite{Haensch02}.   The temperature dependency of the measured visibility is plotted for three separations.  Above $T_C$ (=430nK) the visibility is zero for slit separations $325$nm and $600$nm.  A sudden increase in the visibility occurs below $T_C$ with the maximum visibility being attained close to $T=0$.  The visibility does not vanish above $T_C$ for the slit width $\Delta z = 185$nm as the coherence length at the transition temperature is still larger than the slit width. }\label{fig:Fig5}}
\end{figure}

We can compare our results for spatial coherence to the experimental results.  Let us start by describing in more detail the experiment by H\"ansch {\it et al.}  \cite{Haensch02} that measures the spatial coherence, $\rho_1(\vec{r},\vec{r}\,')$, of a weakly interacting Bose gas. In this double slit type experiment two matter waves were extracted from two regions in the trap separated by $\Delta z$.  The matter waves were allowed to interfere and the degree of spatial cohrence was determined by the the visibility of the fringes.  The position of each `slit' region was determined by the frequency of r.f. field that created transitions between magnetically trapped and untrapped states.  The width of the slits were small, as the r.f. field only coupled weakly to the gas.   The results are given in Fig. \ref{fig:Fig5}  where the spatial coherence is plotted for separations up to $\Delta z = 600$nm.  
 
\begin{figure}[t]\centering
{\includegraphics[width=0.5\textwidth]{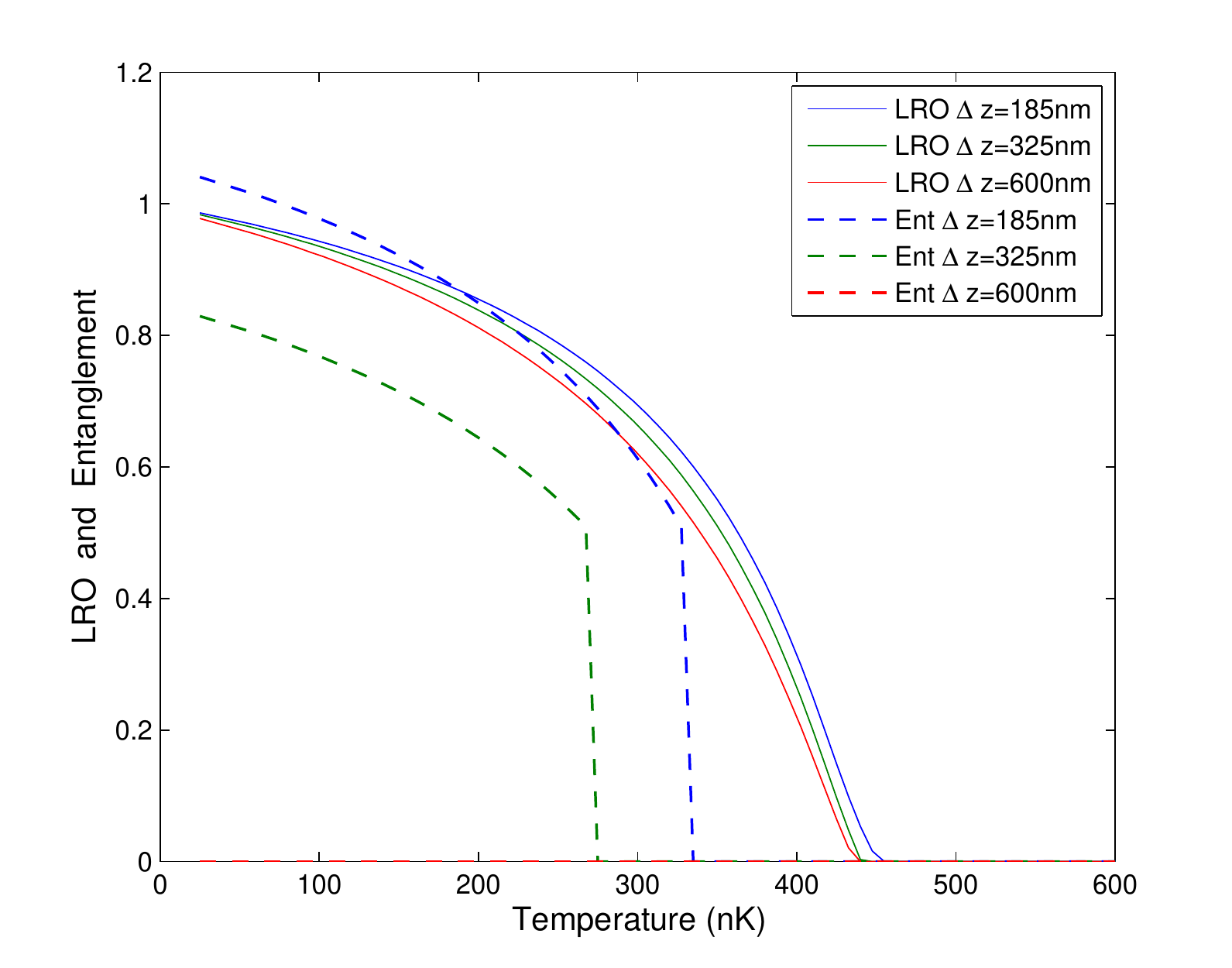}
\caption{The entanglement and spatial coherence at three separations in a non-interacting BEC.  The amount of entanglement between two spatial modes of size $L_A=L_B=350$nm  is  denoted by the dotted lines for slit separations of $\Delta z = 185$nm, $\Delta z = 325$nm and $\Delta z = 600$nm.  The spatial coherence (LRO on legend) as defined by $O_D$, Eq. (\ref{OD}) also for the three same slit separations is is denoted by the solid lines. Spatial coherence is always present when entanglement exists.  The amount of entanglement depends strongly on the separation, and hence population, of the spatial modes as the spatial coherence does not change much between the $\Delta z= 185$nm and $\Delta z = 600$nm.  There is no entanglement for the slit separation $\Delta z = 600$nm because, while there is spatial coherence, the global purity of the two regions is lower than the reduced purity, due to the increased separation of the regions.       Colour online.}\label{fig:Fig7}}
\end{figure}

We can make comparisons between the experimental results measuring spatial coherence between two regions in Fig. \ref{fig:Fig5} and our theoretical results for entanglement and spatial coherence in Fig. \ref{fig:Fig7}.    Let us first take a look at the spatial coherence in both cases.  Spatial coherence defined by $O_D$ Eq. \ref{OD}, is denoted by the solid lines in Fig. \ref{fig:Fig7}.  There is greater spatial coherence between two regions that are closer together.  In our theoretical results the spatial coherence for all three separations is very similar.  If we compare this to the experimental results of Fig. \ref{fig:Fig5}, we see that the spatial coherence for separations $\Delta z=600$nm and $\Delta z=325$nm decreases much faster than our predictions.  This is because Bosons in a real gas interact with each other, which depletes ground state energy level and hence reduces the spatial coherence.  We have not included interactions in our analysis as they only reduce the amount of spatial coherence by a small fraction, which would in turn only effect the amount of entanglement present by a small amount.

The entanglement in Fig. \ref{fig:Fig7} is represented by dashed lines and exists only when spatial coherence exists.   Initially the amount of entanglement between the regions is governed by the behaviour of the one-bodied reduced density matrix, $\rho(\vec{r},\vec{r})$.  When the spatial coherence, $\rho(\vec{r},\vec{r})$, between two regions falls below the probability for the Bosons to not occupy the regions the entanglement falls sharply to zero.  The global purity is larger than the reduced purity to begin with due to the presence of the coherence terms, but as the gas is heated the coherence terms reduce (solid lines Fig. \ref{fig:Fig7}) until the global purity and reduced purity are equal.  The entanglement then disappears. There is no entanglement between the regions when their separation is greater than $\Delta z = 600$nm, even when spatial coherence between the regions is maximum (at $T=0$). This is due to the Bosons gathering around the centre of the trap so that the probability to occupy the regions is very low. Here the state of two regions is very mixed, even at $T=0$.  On the other hand if the regions $A$ and $B$ occupy the entire trap, the amount of entanglement between them depends only on the spatial coherence between $A$ and $B$ and decreases steadily to zero as the spatial coherence goes to zero.

\section{Multi-mode entanglement in a BEC}\label{multientsec}

One must carefully choose the set of spatial modes to ensure they detect the entanglement in the gas. We argue now why a BEC at zero temperature is entangled, even though our choice of spatial modes may not detect the entanglement, such as two regions separated by $\Delta z= 600$nm.

At zero-temperature, $T=0$, all of the Bosons occupy the ground state and the total state of the system, $|\Psi\rangle$, is a pure state.  If the state, $|\Psi\rangle$, were separable with respect to three spatial modes it would be a product state as follows $|\Psi\rangle = |\psi_A\rangle\otimes|\psi_B\rangle\otimes|\psi_C\rangle$.  If region $C$ were then traced out, the resulting state, $|\Psi_{AB}\rangle$, would also be a pure state, $|\Psi_{AB}\rangle=|\psi_A\rangle\otimes|\psi_B\rangle$ and would consequently have unit purity, $\mu = \textrm{tr}[\rho_{AB}^2]=1$.  However in a BEC, the purity, $\mu$, of the state, $\rho_{AB}$, is always less than one, $\mu= \textrm{tr}[\rho_{AB}^2]<1$, unless regions $A$ and $B$ together occupy the entire confining volume.  Therefore, when the purity is less than one, $\mu= \textrm{tr}[\rho_{AB}^2]<1$, we can conclude that the state of three spatial modes, $|\Psi\rangle$, was not a product state to begin with.  This means that a BEC of fixed particle number at zero-temperature is always entangled, because a state of separable regions would have unit purity.  We can also investigate multi-mode spatial entanglement to fully understand how the choice of spatial modes affects the detection of  entanglement in the BEC phase transition.

We will investigate entanglement between three spatial modes, $A$, $B$ and $C$, that together occupy an entire confining volume, $V$, at finite temperatures, $T$.    We can no longer use the purity measure of entanglement as this is only valid for bipartite states.  However, there are suitable entanglement witnesses that detect genuine tripartite entanglement.  An entanglement witness, $\Pi_W$ is an observable that, by our notation, has a positive expectation value for entangled states, $\textrm{tr}[\Pi_W\rho_{ABC}]>0$, and a negative expectation value for separable states.  We will use the witness $\Pi_W=|W_3\rangle\langle W_3|-2/3$, where $|W_3\rangle=1/\sqrt{3}(|100\rangle+|010\rangle+|001\rangle)$, as our single particle state, $\rho_{ABC}$, belongs to the $W$ class of tripartite entangled states.  As we are studying a continuous variable system, we shall write the entanglement witness, $\Pi_{W}$, in terms of field operators, $\hat{\psi}^{\dagger}(\vec{r})$ and $\hat{\psi}(\vec{r})$, as
\begin{equation}
\Pi_W= \sum_{ij=A,B,C}\int_id\vec{r}\,\int_jd\vec{r}\,'g(\vec{r})g^*(\vec{r}\,')\hat{\psi}^{\dagger}(\vec{r})|0\rangle\langle 0 |\hat{\psi}(\vec{r})-2,
\end{equation}
where we have used $|1\rangle_A=\hat{\psi}_A^{\dagger}|0\rangle=\int_Ad\vec{r}\,g(\vec{r})\hat{\psi}^{\dagger}(\vec{r})|0\rangle$.  Here $g(\vec{r})$ takes the role of a detector profile that specifies how we average over the set of points included in the mode.

\begin{figure}[t]\centering
{\includegraphics[width=0.5\textwidth]{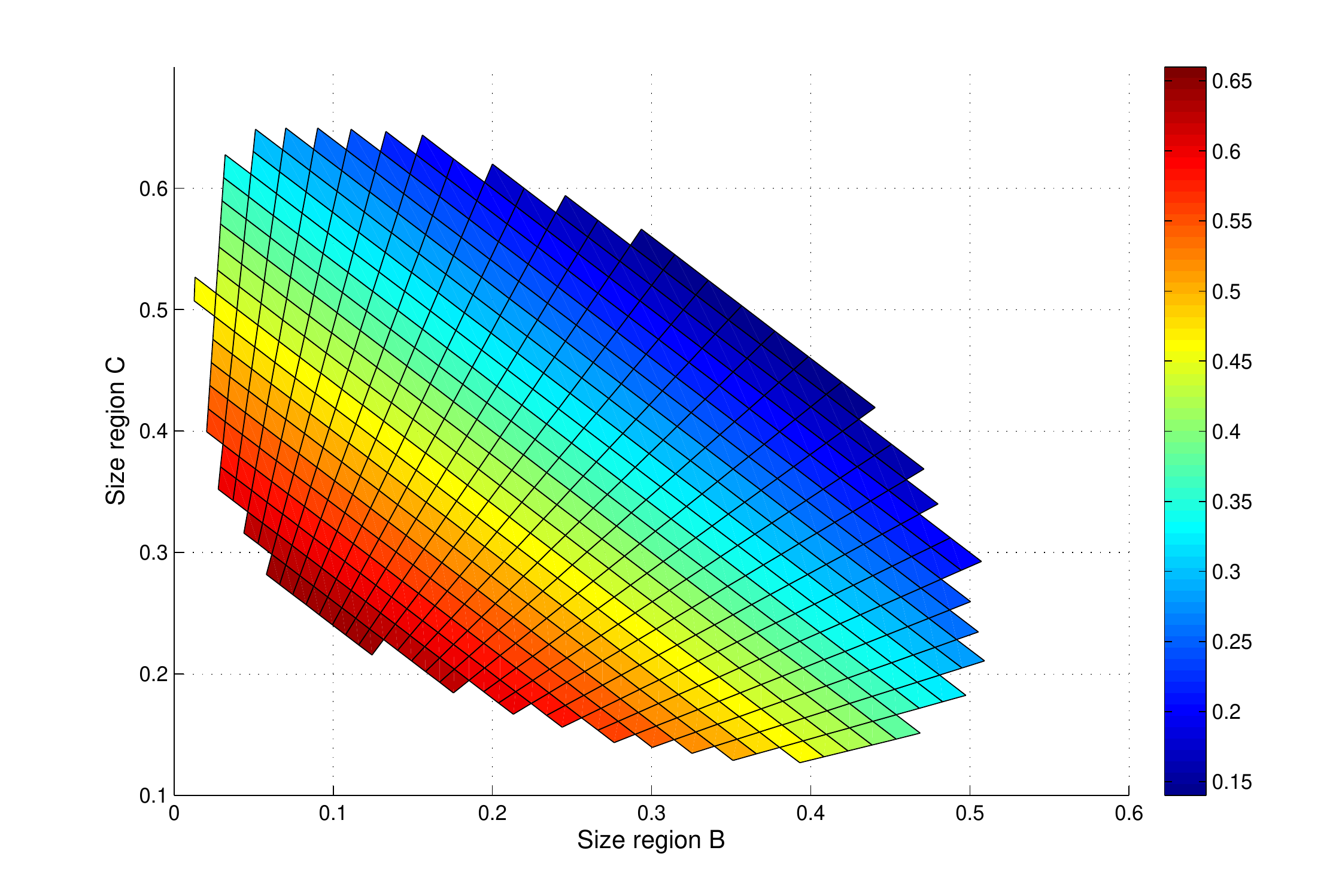}
\caption{Tripartite entanglement, as detected by the entanglement witness $\mathcal{W}$, of a Bosonic gas trapped at zero temperature, $T=0$, in an infinite square well, as a function of the lengths, $L_A, L_B$ and $L_C$, of the spatial modes $A$, $B$ and $C$ respectively.  The lengths are given as fractions of the well length, $L$, and the legend indicates the length of the spatial mode $A$, $L_A$.  Entanglement exists when the probability to occupy the regions are of the same order.  Entanglement does not exist between the three spatial modes, $A$, $B$ and $C$, when any one region is larger than $2/3\,L$.  Region $B$ can become much smaller, $L_B=0.02\,L$, than the other two regions and entanglement still exists.  See text for discussion.   Colour online.}\label{fig:zerotemp}}
\end{figure}
\begin{figure}[t]\centering
{\includegraphics[width=0.5\textwidth]{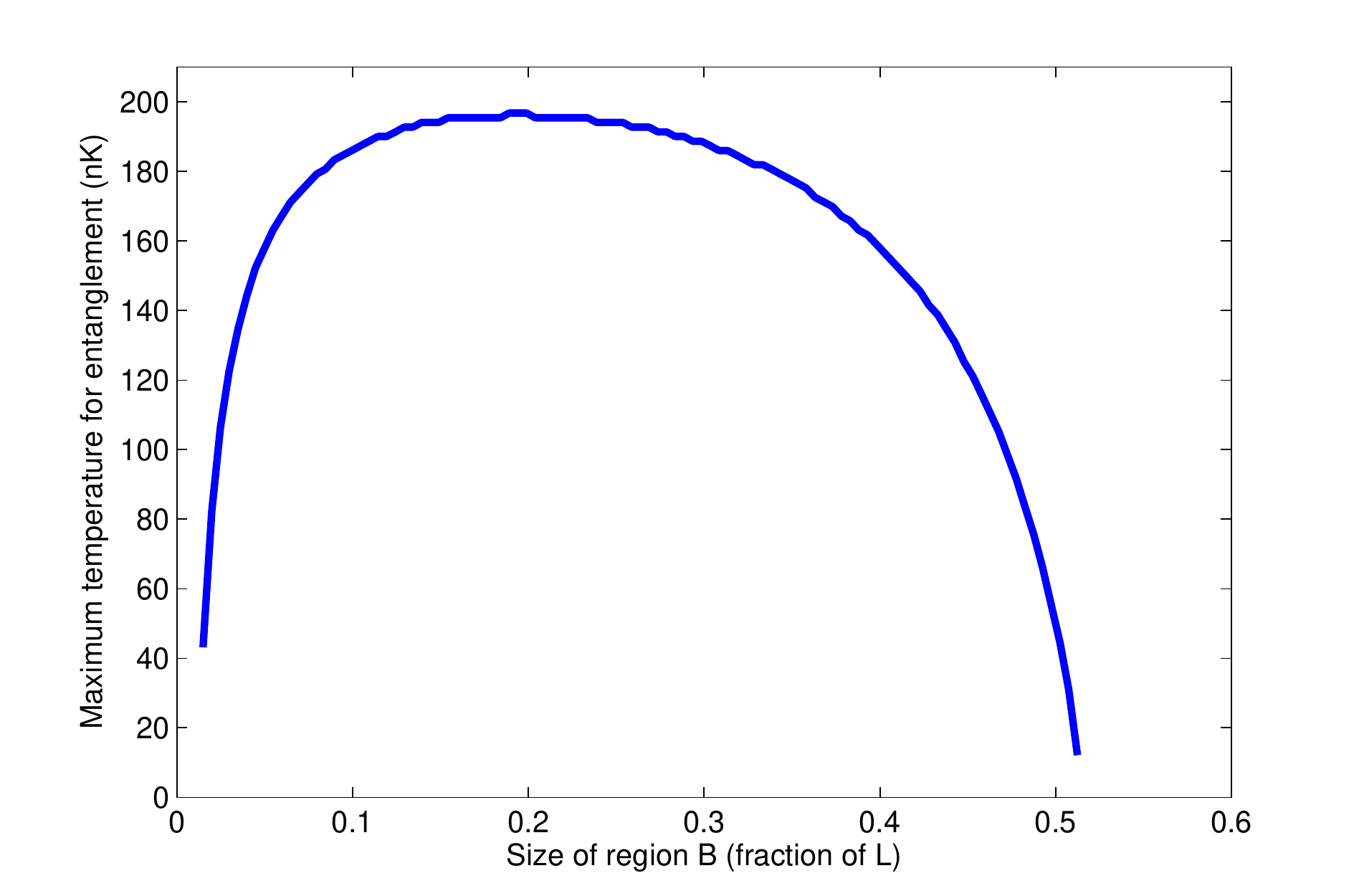}
\caption{The maximum temperature for entanglement, as detected by the entanglement witness $\mathcal{W}$, between three spatial modes of a Bosonic gas in a three-dimensional uniform potential of volume $V=L^3$, as a function of the size of region $B$, $L_B$.  Note that region $B$ is situated between regions $A$ and $C$, whose lengths are taken to be equal, $L_A=L_C$.  The highest temperature for entanglement is $T_{max}=196$nK, which is below the critical temperature for condensation, $T_C=270$nK, which occurs when the probability to occupy all three regions is equal.   See text for further discussion. }\label{fig:maxtempent}}
\end{figure}

We are now in a position to construct an entanglement witness, $\mathcal{W}=\textrm{tr}[\Pi_W\rho_{ABC}]$, that detects tripartite entanglement of a single-particle in the BEC between the three modes.  
The witness, $\mathcal{W}$, is
\begin{equation}
\label{EntWit}
\mathcal{W}=\sum_{ij=A,B,C}\int_id\vec{r}\,\int_jd\vec{r}\,'\,g(\vec{r})g(\vec{r}\,')\rho_1(\vec{r},\vec{r}\,')-2,
\end{equation}
where $\rho_1(\vec{r},\vec{r}\,')=\sum_{\vec{k}}P_{\vec{k}}\phi_{\vec{k}}(\vec{r})\phi^*_{\vec{k}}(\vec{r}\,')$.      When $\mathcal{W}>0$ there is tripartite entanglement between the three spatial modes  $A$, $B$ and $C$.  As we would like to know how entanglement is related to spatial coherence and the one-bodied reduced density matrix $\rho_1(\vec{r},\vec{r}\,')$, it suffices to calculate the entanglement of a single particle in the Bose gas.  This is because while $\rho_1(\vec{r},\vec{r}\,')$ detects entanglement, it does not differentiate between single-particle and $N$-particle entangled states, (see section \ref{BECPhaseTrans}).

The entanglement between three spatial modes, $A$, $B$ and $C$, depends solely on the behaviour of the spatial coherence between each pair of regions, as here we have not traced out any of the gas.  If we consider a state at very high temperature, $T>>1$ every point  in space will be separable from the points immediately next to it, as the de Broglie wavelengths of the atoms become infinitesimally small.  Correspondingly, all of the off-diagonal coherence terms present in the witness $\mathcal{W}$ disappear. The three regions are of course separable, as $\mathcal{W}=-1$. 

\subsection{Zero temperature tripartite entanglement}

On the other hand, at zero temperature, all of the Bosons occupy the ground state and coherences stretch across the entire system.  Whether the three spatial modes are entangled in this case depends only on the relative sizes of the three regions $L_A$, $L_B$ and $L_C$.  We investigate the entanglement between three spatial modes as detected by our entanglement witness,  $\mathcal{W}$, as a function of the region sizes.  We consider again a Bosonic gas of fixed particle number, $N$, trapped at zero-temperature, $T=0$, in the ground state, $\phi_0(r)=\sqrt{2/L}\sin(\pi x/L)$, of a  one dimensional infinite square well of length, $L$. 

Entanglement  exists only when the probabilities to occupy the regions are of the same order.  See Fig. \ref{fig:zerotemp}.   Regions $A$ and $C$ are located on the very left and the very right of the well respectively and region $B$ sits between them.   Entanglement exists when the sizes of regions $A$ and $C$ are between $0.65\,L\leq L_{A,C}\leq0.125\,L$.  Region B, however, can become much smaller, $L_B\approx0.02\,L$ and still be entangled to both regions $A$ and $C$, due to it's position in between regions $A$ and $C$.  This is because the probability for particles to occupy a small region in the centre of the trap is much higher than the probability to occupy the same small region at the edge of the trap. 

To see more clearly why this is the case let us consider a general $W$ state, $|\psi_W\rangle = \alpha|100\rangle+\beta|010\rangle+\gamma|001\rangle$.  Let the coefficient, $\alpha$ become much less than the other two coefficients, $\alpha <<\beta,\gamma$. The first term, whose coefficient is $\alpha$,  can be set to zero.  The first subsystem (denoted by the first position in the kets)   is then no longer entanged to the other two and the $W$-witness, $\Pi_W$,  no longer detects any tripartite entanglement.  This is what, in fact, happens here when one or two of the regions becomes relatively small compared to the remaining one(s).  However, the ground state of the gas is still entangled even when our witness $\mathcal{W}$  does not detect it.  This is because the spatial modes chosen were not the optimum choice, i.e. we could have uncovered entanglement between smaller regions of space by choosing multiple ($n>3$) small spatial modes.  It is an open problem how the amount of entanglement scales with an increasing number of regions.

\subsection{Thermal tripartite entanglement}

In between  high and zero temperatures there is an increase in the thermal de-Broglie wavelength of the atoms.  The atoms spread over neighbouring regions of space and quantum coherences form between them.  At a certain temperature the coherences between the three spatial modes will become big enough so that the regions become entangled.  We will see that entanglement forms below the critical temperature for condensation.  Let us consider a non-interacting Bose gas of $N=30,000$ atoms, confined in a three-dimensional uniform potential of volume, $V=L^3$, where $L=818$nm. The space is divided into three regions, $A$, $C$ and $B$ of lengths $L_A=L_C$ and $L_B$ respectively along one axis and of length $L$ along the other two axes.  Region $B$ is situated in between regions $A$ and $C$.  The atoms occupy the momentum modes $\phi_{\vec{k}}(\vec{r})=\sqrt{2/L}\sin(\vec{k}\vec{r})$. The critical temperature for condensation  is $T_C=270$nK.   

Fig. \ref{fig:maxtempent} shows the maximum temperature, $T_{max}$, for which entanglement exists as a function of the size of region $B$, $L_B$ (which is also the separation of regions $A$ and $C$), as determined by the entanglement witness, $\mathcal{W}$, Eq. (\ref{EntWit}).  The warmest temperature that entanglement survives at is $T_{max} = 196$nK, which corresponds to the region sizes $L_A=L_C=0.4\,L$ and $L_B=0.2\,L$. At this point there is approximately equal probability, $P$, for the Bosons to occupy any region, i.e. $P_A=P_C\approx P_B$.  At zero temperature this spatial configuration of modes, $P_A\approx P_C\approx P_B$, represents a state with the maximum amount of possible entanglement, as maximally entangled states usually posses equal probability amplitudes. I.e. the maximally entangled Bell states all have probability amplitudes $a=1/\sqrt{2}$.  As the temperature increases from zero Kelvin,  a fraction of the Bosons occupy higher momentum modes and the entangled state of three spatial modes becomes increasingly mixed.  At some temperature, these  quantum states become so mixed that they are no longer entangled.  States with only a small amount of entanglement at $T=0$ need only a small amount of extra mixedness to become separable so are destroyed by only a small amount of heat.  In contrast consider the entangled states with some maximum amount of entanglement at $T=0$. The entanglement in these states survives to higher temperatures as they tolerate more mixing before they become separable.

As the number of Bosons, $N$, in the gas increases the critical temperature for condensation, $T_C$ also increases for a fixed volume.  How does the maximum temperature for single-particle  entanglement, $T_{max}$, change as $N$ increases?  Naturally, $T_{max}$ increases with $N$ and $T_C$, but the ratio, $T_{max}/T_C$ also tends to unity, $T_{max}/T_C\rightarrow 1$.  Thus, as the number of Bosons in the condensate, $N$, becomes very large, $N\rightarrow\infty$, the maximum temperature for entanglement $T_{max}$ tends to the critical temperature for condensation, $T_{max}\rightarrow T_C$.  In this limit, and for the optimum region configuration described above i.e. when $P_A=P_B=P_C$, spatial entanglement and spatial coherence become equivalent.  In other words entanglement between three regions of space and spatial coherence between the same regions both exist for an identical set of parameters.  
\section{Discussion and conclusion}\label{concludesec}

Let us summarise the results of this paper.  Initially we used a purity measure of entanglement to derive an entanglement witness, $\mathcal{E}$, Eq. (\ref{entinequality}), that indicated whether two spatial modes of a non-interacting Bosonic gas of fixed particle number were entangled.  The witness, $\mathcal{E}$, is parameterised by the one-body reduced density matrix, $\rho(\vec{r},\vec{r}\,')$, that  measures the spatial coherence between points $\vec{r}$ and $\vec{r}\,'$, and the probability to occupy one of the regions $p$.   When the two regions occupy the entire confining geometry ($p=1/2$) the entanglement witness, $\mathcal{E}$, depends only on the one-body density matrix and entanglement and spatial coherence are equivalent.   If the probability to occupy a region is less than a half, $p<1/2$, then spatial coherence between the regions is only necessary for entanglement.   

We plotted the results for the case of a cigar-shaped harmonic potential and as expected entanglement only exists below the critical temperature for condensation, $T_C$, when there is coherence between two separated regions.  The amount of entanglement increases when the spatial modes are closer together because the population is located at the centre of the trap.  If there were an equal probability to find Bosons at every point in the trap there would be an equal amount of entanglement between all regions of space.  The non-local quantum part of the one-bodied reduced density matrix is as expected very long ranged.
 
To gain a greater understanding of entanglement in the BEC phase transition it was also useful to consider multi-mode entanglement, as three or more spatial modes can still be entangled even when two spatial modes are not.  We derived a second entanglement witness, $\mathcal{W}$, to detect entanglement between three spatial modes using an existing entanglement witness, $\Pi_W$.  In this case the three regions occupied the entire confining volume so the witness, $\mathcal{W}$ was parameterised exclusively by the one-body reduced density matrix.  The probability to occupy all three regions must be of the same order for entanglement to exist.  At zero temperature, $T=0$, the state with maximum entanglement corresponds to the region configuration that allows the probabilities for occupation of each region to be equal, $P_A=P_B=P_C$.  The entanglement in this state was also more robust to mixing and survived until the highest temperatures.

In the introduction we spoke about two conjectures concerning the entanglement in quantum phase transitions.  Namely that entanglement is responsible for the long-range correlations that develop in the transverse Ising and XY models at the quantum critical point and that at such critical points the system is maximally, or {\it critically}, entangled \cite{Osborne02}.   We will discuss these conjectures in the context of the BEC phase transition.  Even though the phase transitions are very different in nature, our results may act as a guide for how entanglement behaves in other such critical systems.  

In contrast to a quantum phase transition,  the BEC phase transition is classical and is driven by thermal fluctuations.  One could, however, induce the BEC phase transition at zero temperature by increasing the density of Bosons until their wavefunctions start to overlap, and a BEC was formed.  It is therefore sensible to compare the entanglement properties of  BECs at zero temperature to entanglement in quantum phase transitions. Another difference between entanglement in BECs and entanglement in discrete lattice type systems is the couplings between sites.  In discrete systems one normally considers nearest neighbour couplings whereas in a BEC there are pairwise couplings between all regions.  The entanglement structure is far richer in a BEC, with entanglement stretching well beyond neighbouring regions.  

Entanglement below the critical temperature for condensation occurs due to the long-range correlations that distribute Bosons coherently over the spatial modes.  We have seen that spatial coherence is a precursor to entanglement and we know, of course, that the entanglement does not cause the spatial correlations.  When the probability for Bosons to occupy each region is equal, entanglement and spatial coherence between the regions are equivalent and the system has the maximum possible amount of entanglement. In discrete systems, where the  sites are pre-defined, could entanglement and spatial coherence also be equivalent?  At criticality long-range correlations do occur throughout the system, but one would need a suitable multi-partite entanglement witness to fully characterise the behaviour of entanglement between all the sites at that point.  

Let us finally discuss two possible extensions of this work.  Firstly one could characterise the behaviour of entanglement between $M$ spatial modes.  We know that genuine multi-mode entanglement only exists below the critical temperature for condensation, due to the requirement of spatial coherence between distant regions. However, it is interesting to ask whether there could be entanglement above the critical temperature for condensation. Consider regions smaller than the Boson's de Broglie  wavelength, so that there was spatial coherence between neighbouring modes, entanglement of the whole system could then be built up from bipartite entanglement, to form some sort of entangled graph state.

A second reason for understanding how spatial coherence is related to entanglement lies in the fact that it is becoming increasingly accepted that BEC like coherences can occur in systems at higher temperatures.  In this case the systems interacts strongly with the environment and a constant supply of energy lift the systems modes out of the Planck distribution causing the strong excitation of a single mode \cite{Hartmann07}.  This is particularly interesting when one realises that biological systems may show such behaviour where quantum coherence is thought to be present in energy transfer in photosynthetic systems \cite{Engel07}.

\medskip

The author acknowledges the support of the Engineering and Physical Sciences Research Council in UK for funding. She would like to thank Vlatko Vedral, Caslav Brukner, Johannes Kofler and Sandu Popescu for useful discussions about this work and Bruno Sanguinetti for unparalleled assistance with her Mac.

\end{document}